\documentclass[twocolumn, aps,prl, superscriptaddress,showpacs]{revtex4}

\usepackage{mathptmx}
\usepackage{subfigure}
\usepackage{dcolumn}
\usepackage{amsmath,amssymb}
\usepackage{bm}
\usepackage{color}
\usepackage{latexsym}
\usepackage[english]{babel}
\usepackage{times}

\usepackage{psfrag,graphicx} 
\usepackage{epsf}  
\usepackage{amsfonts}
\usepackage{natbib}
\usepackage{epstopdf}
\usepackage{appendix}
\usepackage[colorlinks=true,citecolor=blue,linkcolor=blue]{hyperref}

\begin{document}

\title{Simulating $Z_2$ topological insulators with cold atoms in a one-dimensional
optical lattice}
\author{Feng Mei}
\affiliation{Laboratory of Photonic Information Technology, LQIT $\&$ SIPSE, South China Normal University, Guangzhou 510006, China}
\affiliation{Centre for Quantum Technologies and Department of Physics, National University of Singapore, 3 Science Drive 2, Singapore 117543, Singapore}

\author{ Shi-Liang Zhu}
\affiliation{Laboratory of Quantum Information Technology and SPTE, South China
Normal University, Guangzhou, China}

\author{Zhi-Ming Zhang}
\email{zmzhangATscnu.edu.cn}
\affiliation{Laboratory of Photonic Information Technology, LQIT $\&$ SIPSE, South China Normal University, Guangzhou 510006, China}

\author{C. H. Oh}
\email{phyohchATnus.edu.sg}
\affiliation{Centre for Quantum Technologies and Department of Physics, National University of Singapore, 3 Science Drive 2, Singapore 117543, Singapore}

\author{N. Goldman}
\email{ngoldmanATulb.ac.be}
\affiliation{Center for Nonlinear Phenomena and Complex Systems - Universit\'e Libre de Bruxelles , 231, Campus Plaine, B-1050 Brussels, Belgium}

\date{\today }

\begin{abstract}
We propose an experimental scheme to simulate and detect the properties of time-reversal invariant topological
insulators, using cold atoms trapped in one-dimensional bichromatic optical
lattices. This system is described by a one-dimensional 
Aubry-Andre model with an additional SU(2) gauge structure, which captures the essential properties of a two-dimensional $Z_2$ topological insulator. We demonstrate that 
topologically protected edge states, with opposite spin orientations,  can be pumped across
the lattice by  sweeping a laser phase adiabatically. This process constitutes an elegant way to transfer topologically protected quantum states in a highly controllable environment. We discuss how density measurements could provide clear signatures of the topological phases emanating from our one-dimensional system.
\end{abstract}

\pacs{37.10.Jk, 73.43.-f, 67.85.Lm}
\maketitle

%\section{Introduction}

The quantum Hall (QH) effect, discovered in 1980, provided the first example of
a quantum phase that has no spontaneously broken symmetry. Besides, its universal character and remarkable robustness have been shown to be related to the existence of topological invariants \cite{TI1, TI2, TI3}. The recent discovery of
the quantum spin Hall (QSH) effect, in materials displaying strong spin-orbit coupling, has opened the path for a new family of topological states: the Z$_{2}$ topological insulators  {\color{blue}\cite{TI1,TI2}}. Since then, the search for
topological phases of matter has become a forefront topic in condensed matter
physics \cite{TI3}. In general, topological insulators are insulating in the bulk but they feature
gapless edge or surface states at their boundary. These edge modes are very
robust and therefore persist in the presence of impurities. The delicate
control over these edge modes has attracted considerable interest for the
realization of quantum spintronic and magnetoelectric devices {\color{blue}%
\cite{TI3}}. Furthermore, in proximity of superconductors, topological insulators lead to non-Abelian excitations that could
lead to a new architecture for topological quantum computation {\color{blue}%
\cite{TQC}}.

Nowadays, cold atoms trapped in optical lattices are widely recognized as powerful experimental tools to mimic a wide range of systems originally stemming from condensed matter physics {\color{blue}\cite{QS1,QS2}}. Recently, many experimental
efforts have been focused on the experimental realization of synthetic magnetic fields and spin-orbit coupling for ultracold atoms \cite{dalibard,GF1,GF2,GF3,GF4}, which set the stage for
the simulation of topological insulators and fractional quantum Hall states. In particular, several proposals have been
suggested to realize the QH and QSH states
using these technologies {\color{blue}\cite{AH,SH}}. 

The QH and QSH phases are realized in 2D systems subjected, respectively, to strong magnetic or spin-orbit couplings.  Surprisingly, several properties associated to these topological states can already be probed through a one-dimensional reduction of these systems. This idea has  been explored theoretically and experimentally in 1D quasicrystals {\color{blue}\cite{TP}}, which reproduce the Hofstadter-Aubry-Andre QH model \cite{HP,AA}. This elegant discovery opens the possibility to investigate QH physics using one-dimensional optical lattices {\color{blue}\cite{1QH}}.

In this paper, we propose an experimental scheme to simulate and detect $Z_2$ topological states with cold atoms trapped in a 1D optical lattice. We show that these topological phases can be described by a generalized 1D Harper equation with an additional SU(2) gauge structure, which could be simulated with a two-component atomic gas trapped in a 1D bichromatic optical lattice. By adjusting the corresponding laser configuration, one is able to probe several properties of $Z_2$ topological states. In particular, one could transfer the spin-resolved edge states from one edge to the other, and measure these states through density measurements. One also discusses the possibility to define a  $Z_2$ topological invariant in this 1D framework, allowing to distinguish between  trivial and non-trivial topological states. Finally, we describe how density measurements could provide an efficient tool to measure these invariants in the present context.

Let us start by presenting a specific 2D tight-binding model, which has been introduced in Ref. {\color{blue}\cite{SH}} to simulate a $Z_2$ topological insulator with two-component fermions in an optical square lattice. The corresponding second-quantized Hamiltonian reads
\begin{eqnarray}
H &=&t\sum\limits_{m,n}c_{m+1,n}^{\dagger}e^{i\theta
_{x}}c_{m,n}+c_{m,n+1}^{\dagger}e^{i\theta _{y}(m)}c_{m,n}+h.c \notag \\
&&+\lambda_{\text{stag}} \sum\limits_{m,n}(-1)^{m}c_{m,n}^{\dagger}c_{m,n},\label{hamilton}
\end{eqnarray}%
where $c_{m,n}$ is a two-component (spin 1/2) field operator defined on the lattice site ($x=ma$, $y=na$), $a=1$ is the lattice spacing, and $t$ is the nearest-neighbor hopping amplitude. Here, the spin-1/2 structure derives from the fact that each site hosts atoms in two  internal states \cite{GF4}. The second line of Eq. \eqref{hamilton} describes an on-site staggered potential with amplitude $ \lambda_{\text{stag}} $, along the $x$ direction, which has been introduced to drive transitions between different topological phases. The Peierls phases $\theta _{x}$ and $\theta _{y}(m)$, which accompany the hopping along the $x$ and $y$ directions, are engineered within this tight-binding model to simulate the analog of spin-orbit couplings, and are expressed in terms of the Pauli matrices $\sigma_{x,y,z}$. The space-dependent operator $\theta _{y}(m)=2\pi m \alpha \sigma _{z}$ reproduces the effect of the intrinsic spin-orbit coupling \cite{KaneMele}: it corresponds to opposite ``magnetic" fluxes $\pm \alpha $ for each spin component and generates QSH phases. The constant operator $\theta _{x}=2\pi \gamma \sigma _{x}$ corresponds to a spin-mixing perturbation, and thus simulates a Rashba spin-orbit coupling term {\color{blue}\cite{KaneMele}}.

The Hamiltonian \eqref{hamilton} being translationally invariant along the $y$ direction, one can explore its properties by imposing periodic boundary conditions along this direction. Considering this cylindrical geometry, the single-particle wave function is expressed as $\psi(m,n)= \exp (i k_y n) \Psi (m)$, where $k_y$ is the quasimomentum along the periodic
coordinate, and where the two-component wave function $\Psi _{m}=\left( \Psi _{m}^{\uparrow },\Psi _{m}^{\downarrow }\right)
^{T}$ satisfies a generalized 1D Harper equation \cite{HP,AA}
\begin{eqnarray}
E\Psi _{m}(k_y) &=&t(e^{i\theta _{x}}\Psi _{m+1}(k_y)+e^{-i\theta _{x}}\Psi
_{m-1}(k_y)) \notag \\
&&+R(m,k_y)\Psi _{m}(k_y).\label{harper}
\end{eqnarray}%
We have introduced the onsite $2 \times 2$ matrix 
\begin{align}
R(m,k_y)&=2t \text{diag}(\cos (2\pi \alpha
m+k_y),\cos (2\pi \alpha m-k_y)) \notag \\
&+\lambda_{\text{stag}} (-1)^{m} I,
\end{align}
where $I$ is the identity matrix.  The Harper Eq. \eqref{harper} therefore describes the dimensional reduction of the initial system \eqref{hamilton}, but still captures the essential properties of its topological phases.  The energy spectrum $E=E(k_y)$ obtained by solving Eq. \eqref{harper} displays several energy bands that describe the bulk, but also several gapless states which constitute clear signatures of QSH topological phases \cite{KaneMele,SH}. We note that Eq. \eqref{harper} generalizes the spinless Harper equation obtained by Aubry-Andre, in their study of Anderson localization {\color{blue}\cite{AA}}. Recently, the Aubry-Andre model has been simulated in a 1D bichromatic optical lattice {\color{blue}\cite{AL}}, which motivates us to propose a generalization of their setup to investigate the physics stemming from our spin-1/2 Eq. \eqref{harper}. 

Let us first focus on the case $\gamma=\lambda_{\text{stag}}=0$, where the spin-mixing perturbation $\theta_x$ and the staggered potential are absent. The realization and implication of these terms will be discussed at the end of this work. In this case, the spin-1/2 model described by Eq. \eqref{harper} can be realized by trapping a two-component atomic gas in a primary 1D optical lattice $V_1(x)=V_1 \cos ^2 (k_1 x)$, with wave number $k_1$. In this configuration, and for a sufficiently deep potential, the atomic system is governed by the tight-binding Hamiltonian
\begin{equation}
H_1 =t\sum_{m} c_{m+1}^{\dagger} c_{m} +c_{m-1}^{\dagger} c_{m},
\end{equation}
where $c_{m}$ is a two-component (spin 1/2) field operator defined on the lattice site ($x=ma$) and $m=1 , \dots , L$. Then, the onsite term characterized by the $2 \times 2$ matrix $R(m)$ can be realized by two  weak state-dependent lattices, with wave number $k_2$, which act independently on the two atomic states \cite{spindep}. For our purpose, the optical potentials associated to these state-dependent lattices have the form $V_{\uparrow, \downarrow}(x)=V_s \cos ^2 (k_2 x \pm \phi/2)$, which can be produced by two counterpropagating laser beams with linear polarization vectors forming an angle $\phi$ \cite{spindep2,spindep3,QS1}. These additional lattices interfere with the primary lattice, and supposing that $V_s \ll V_1$, simply lead to the onsite perturbation term \cite{AL,bichro,bichro2}
\begin{equation}
H_{\text{2}}= \Lambda  \sum_{m} c^{ \dagger}_{m \uparrow} c_{m \uparrow} \cos(2 \pi \beta m +\phi) +  c_{m \downarrow}^{\dagger} c_{m \downarrow} \cos(2 \pi \beta m -\phi),
\end{equation}
where $\Lambda \sim t \ll V_1$, $\beta=k_{2}/k_1$. Although the phase $\phi$ could be affected by an overall shift of $V_1(x)$ with respect to $V_{\uparrow, \downarrow}(x)$, this parameter can be monitored through various technics (cf. the experimental methods in Ref.  \cite{bichro2}, but also the studies of Refs. \cite{larcher} on the effects of uncontrolled phases in cold-atom Aubry-Andre models). Besides, we note that the topological properties described in this work remain constant for small variations of the parameter $\phi$ (cf. below). In this configuration, the single-particle equation associated to the total Hamiltonian $H_{\text{tot}}=H_1+H_2$ reads
\begin{equation}
E \Psi _{m}(\phi ) =t \, (\Psi _{m+1}(\phi )+\Psi _{m-1}(\phi )) +S(m,\phi)\Psi _{m}(\phi ), \label{simpleq}
\end{equation}
where $S(m,\phi)=\Lambda\text{diag} (\cos (2\pi \beta m+\phi ),\cos (2\pi \beta m-\phi
))$. Therefore, a direct mapping from this 1D Aubry-Andre system to the 2D setup of Ref. {\color{blue}\cite{SH}} is obtained by associating the commensurability parameter $\beta$ to the ``magnetic" flux $\alpha$, the potential strength $\Lambda$ to twice the tunneling amplitude $2 t$, and the phase $\phi$ to the quasimomentum $k_y$. In other words, our spin-1/2 generalization of the Aubry-Andre model, which could be simulated using 1D state-dependent lattices \cite{AL}, could already reveal the topological properties emanating from Eqs. \eqref{hamilton}-\eqref{harper}. It is worth emphasizing that, altough a direct mapping exists between the energy spectra $E(k_y)$ and $E (\phi)$, that respectively correspond to the 2D model of Ref. {\color{blue}\cite{SH}} and the present Aubry-Andre-type model, the parameter $\phi$ is \emph{fixed} in the latter experimental scheme. Therefore, the energy ``bands" depicted by the spectrum $E(\phi)$ involves the \emph{union} of different configurations of the system, obtained by continuously varying the parameter $\phi$. 

Let us investigate the spectral properties of Eq. \eqref{simpleq}: for a fixed value of the parameters $\phi$ and $\Lambda$ (in the following $\Lambda=2t =2$ and $\phi \in [0 , 2 \pi]$), and for a rational value of the commensurability parameter $\beta=p/q$, the spectrum splits into $q$ continua of states (cf. Fig. 1 (a)). These states are delocalized and describe the bulk of our 1D system.  In the example illustrated in Fig. 1 (a), one has set $\beta=1/3$, which leads to three ``bulk subbands". Between these continua of bulk states, and within certain ranges of the parameter $\phi$, one finds two degenerate states with opposite spin, whose amplitudes are localized at the two edges of the system (cf. the states in Fig. 1 (a), which are highlighted by a dot (resp. a star)  for $\phi=1$ (resp. $\phi=2 \pi -1$)). As depicted in Fig. 1 (b),  the spin orientation of these edge states are opposite at the two edges: when $\phi=1$, one finds a spin-up (resp. down) state at $E \approx -t$,  which is localized at $m=L$ (resp. $m=1$). The opposite situation, i.e. spin up (resp. down) at $m=1$ (resp. $m=L$), occurs at the same energy by setting $\phi=2 \pi -1$ (cf. Fig. 1 (b)). 

In the analogous 2D system, these two pairs of states constitute \emph{helical} edge states, which is a hallmark of the QSH effect \cite{KaneMele}: when the Fermi energy is set in the first bulk gap $E_{\text{Fermi}}\approx - t$, each edge is populated by opposite spins traveling in opposite directions. In contrast, in the present context of a 1D lattice, the parameter $\phi$ is fixed, and therefore, each edge is populated by a \emph{single} spin species. However, by adiabatically varying the phase $\phi$ between $[0, 2 \pi]$, following a gapless edge state within the lowest ``bulk gap", one drives an interesting transition between a non-trivial edge-state configuration (e.g. spin up at $m=1$ and spin down at $m=L$) to the opposite configuration (e.g. spin up at $m=L$ and spin down at $m=1$). While the total charge is conserved at each edge, this exquisite process leads to  \emph{spin pumping}. We note that the edge states remain localized during the whole process, except at singular points (e.g. $\phi=\pi$ in the example presented in Fig. 1), where these states connect to the bulk. Therefore, the edge states survive for small variations of the parameter $\phi$, a fact which is in agreement with the topological argument discussed below. 

\begin{figure}[tbp]
\includegraphics[width=9cm]{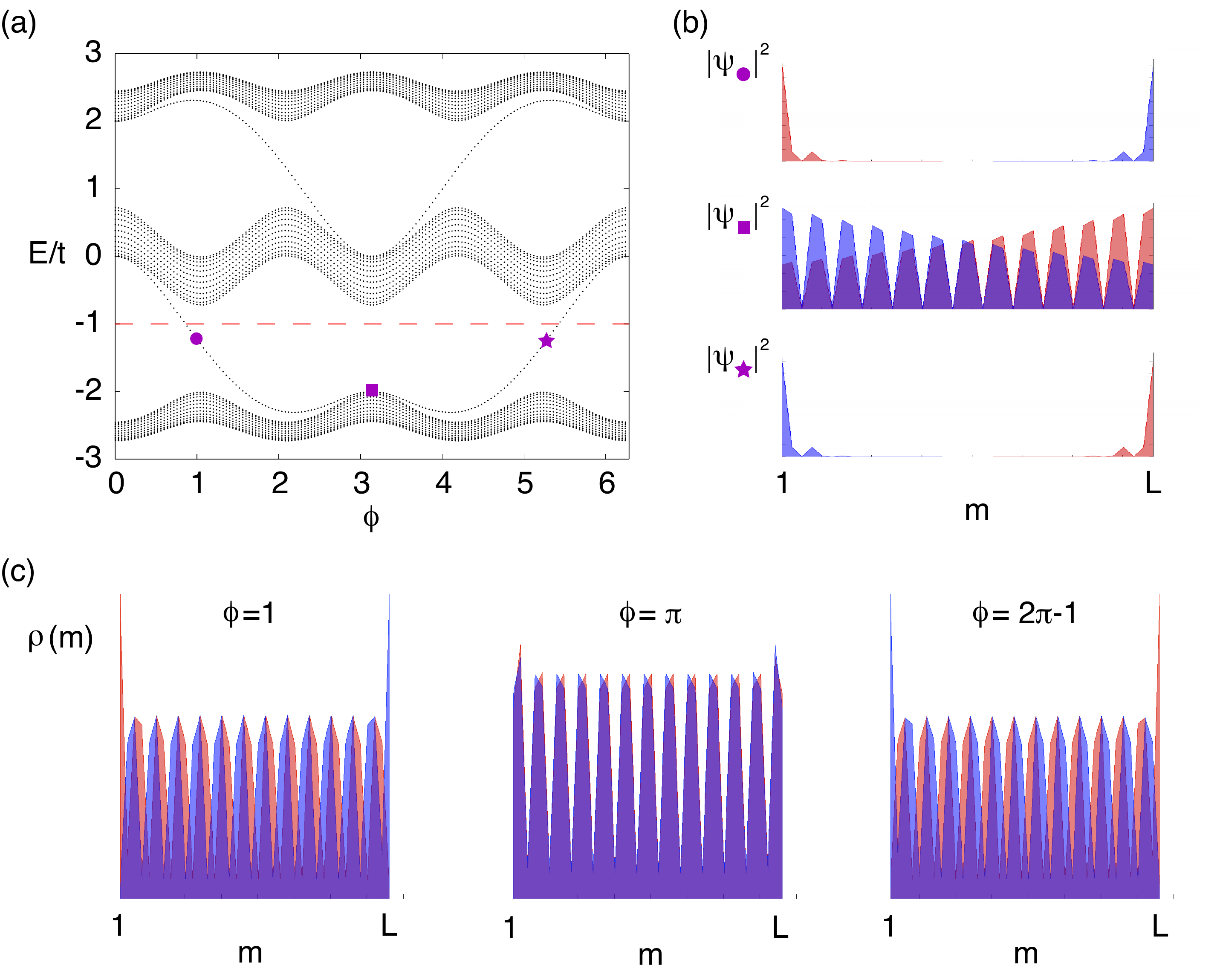}
\caption{(Color online) (a) Energy spectrum as a function of the phase $\phi $ for a 1D lattice with $L=38$ sites. Here $\beta=1/3$ and $\Lambda=2t$. (b) The amplitudes $\vert \Psi_{m \uparrow} \vert ^2$ (resp. $\vert \Psi_{m \downarrow} \vert ^2$) are represented in blue (resp red) as a function of the site index $m$, and correspond to the three states highlighted in (a) at $\phi=1, \pi, 2 \pi -1$. (c) The spin-resolved particle densities $\rho_{\uparrow} (m)$ and $\rho_{\downarrow} (m)$, depicted in blue and red respectively, for $E_{\text{Fermi}}=- t$ and $\phi=1, \pi, 2 \pi -1$. These densities have been computed for infinitely sharp boundaries. Note the inversion of the spin structure at the opposite edges, as $\phi$ is varied.}
\end{figure}

In standard two-dimensional QSH systems, the existence of gapless helical edge states inside a bulk gap is guaranteed by a topological invariant, the so-called $Z_2$ index ($\nu=0,1$), which can be evaluated from the bulk states \cite{KaneMele}. When the Fermi energy is fixed in a bulk gap characterized by the non-trivial value $\nu=1$, an odd number of helical edge state pairs are located at each edge, in which case the system realizes the QSH effect. As long as the bulk gap remains open, the index $\nu$ remains constant, which guarantees the robustness of the edge states against external perturbations. Now, if one maps the example presented in Fig. 1 (a)-(b) to the analogous 2D system (i.e. $\phi \rightarrow k_y$, where $k_y$ takes all the values within the range $k_y \in [0, 2 \pi]$), one observes that each bulk gap of the spectrum $E(k_y)$ hosts two pairs of helical edge states (i.e. one pair for each edge). Therefore, the bulk gaps presented in Fig. 1 would both correspond to the $Z_2$ index $\nu=1$ in a 2D realization of the system \cite{SH}. Mapping this example back to our 1D context (i.e. $k_y \rightarrow \phi$, with $\phi$ fixed), this result indicates that, as long as the bulk gap remains open, there will always be a range  $\phi \in [\phi_1 , \phi_2]$ between which such edge states will be detected.  

Obviously, this topological argument is based on the analogy between our 1D system and its analogous two-dimensional QSH system, whose dimensionality allows to properly define the topological invariant $\nu$ \cite{KaneMele}. However, one can show that this topological invariant can also be rigorously defined in the 1D framework. Indeed, when $\gamma =0$, the $Z_2$ index is simply related to the spin Chern number $\nu = \text{SChN} \text{mod 2}$, where $\text{SChN}=(\text{ChN}_{\uparrow}-\text{ChN}_{\downarrow})/2$ and where $\text{ChN}_{\uparrow, \downarrow}$ are the Chern numbers associated to the up and down spin respectively \cite{SC}. When $\gamma=0$, the spin components are decoupled and the Chern numbers  $\text{ChN}_{\uparrow, \downarrow}$ can be evaluated individually from the standard TKNN expression \cite{Kohmoto1985}
\begin{equation}
\text{ChN}_{\uparrow}=\frac{1}{2 \pi i} \sum_{E_{\lambda} \le E_{\text{Fermi}}} \int_{\mathbb{T}^2} dk_x d \phi \, \mathcal{F} (\vert \Psi_{\uparrow \lambda} (k_x,\phi) \rangle),\label{chern}
\end{equation}
where $\mathcal{F} $ is the Berry curvature associated to the single-particle state $\vert \Psi_{\uparrow \lambda} (k_x,\phi) \rangle$, which is caracterized by the band index $\lambda$ situated below the Fermi energy $E_{\text{Fermi}}$, and where $k_x$ is the quasimomentum. In this expression, one supposes that the parameter $\phi$ evolves continuously along the interval $[0 , 2 \pi]$:  namely, the definition of the Chern number \eqref{chern} requires the union of all the Hamiltonian operators $H(\phi)$. In fact, it was recently shown that such a  Chern number could be rigorously defined for each $\phi$  \cite{TP}, and that it remains constant for all $\phi \in [0 , 2 \pi]$. This result, which is in agreement with the argument based on the 2D analogy (cf. above), guarantees the existence of edge states for certain ranges of the parameter $\phi$ \cite{TP}. Finally, we stress that for the general situation, where spin-mixing is present $\gamma \ne 0$, the topological $Z_2$ index and the SChN are no longer expressed in terms of the individual Chern numbers $\text{ChN}_{\uparrow, \downarrow}$. However, their values obtained in the limit $\gamma \rightarrow 0$ remain constant for finite $\gamma$, as long as their associated bulk gap remains open \cite{SC}. We note that, for the example illustrated in Fig. 1 (a), the Chern numbers are given by $\text{ChN}_{\uparrow, \downarrow}=\pm 1$ in the first gap and $\text{ChN}_{\uparrow, \downarrow}=\mp 1$ in the second, which indeed leads to the non-trivial $Z_2$ index $\nu=1$ in both cases. 

One stresses that the presence of spin-polarized edge states, for a fixed value of the parameter $\phi$, does not necessarily mean that the $Z_2$ index is non-trivial ($\nu=1$): indeed, one should count the number of such states within the whole range $\phi \in [0, 2 \pi]$  to determine whether the number of such helical edge state pairs is odd ($\nu=1$) or even ($\nu=0$) at each edge. This counting procedure, which could be performed experimentally by continuously varying the phase $\phi$, allows to rigorously classify the phases of our 1D model in terms of the $Z_2$ topological index. To illustrate a configuration displaying trivial and non-trivial $Z_2$ phases, one has computed the energy spectrum and edge state structures for the case $\beta=1/6$ (cf. Fig. 2). Here, the first bulk gap is characterized by a non-trivial $Z_2$ index $\nu=1$, as for the previous example discussed above for $\beta=1/3$. However, in the second bulk gap around $E \approx -t$, one would observe two helical edge state pairs at each edge, by varying $\phi \in [0, 2 \pi]$ (cf. Fig. 2 (b)): the second gap is therefore associated to the trivial phase $\nu=0$. This latter result is in agreement with the value $\text{SChN}=2$, which can be computed from Eq. \eqref{chern} for $E_{\text{Fermi}}=-t$ and $\beta=1/6$.

\begin{figure}[tbp]
\includegraphics[width=9cm]{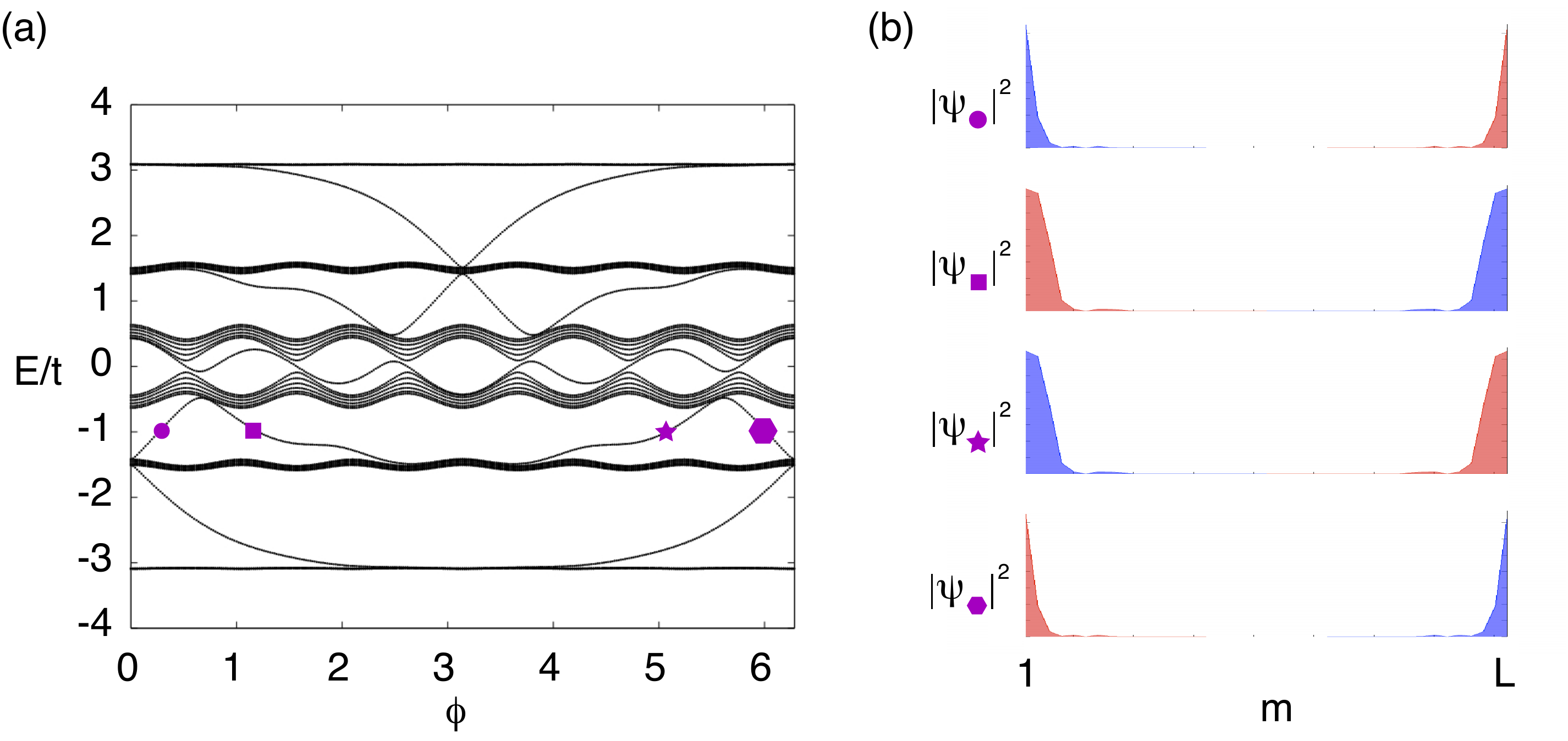}
\caption{(Color online) (a) Energy spectrum as a function of the phase $\phi $ for a 1D lattice with $L=41$ sites. Here $\beta=1/6$ and $\Lambda=2t$. (b) The amplitudes $\vert \Psi_{m \uparrow} \vert ^2$ (resp. $\vert \Psi_{m \downarrow} \vert ^2$) are represented in blue (resp. red) as a function of the site index $m$, and correspond to the four states highlighted in (a), i.e. the states labeled by a dot, a rectangle, a star and a hexagon respectively.}
\end{figure}

Now, let us show how the detection of $Z_2$ topological phases could already be obtained from density measurements. The first strategy would be to directly detect the edge states, which is a realistic task for systems featuring infinitely sharp boundaries and low Fermi energy. Indeed, in this configuration, the edge states will contribute to the particle density in a detectable way. We illustrate in Fig. 1 (c) the spin densities, defined by
\begin{equation}
\rho_{\uparrow , \downarrow} (m)= \sum_{E_{\lambda} \le E_{\text{Fermi}}} \vert \Psi_{\uparrow , \downarrow, \lambda} (m) \vert^2,
\end{equation}
for $\phi= 1, \pi$ and $2 \pi -1$ and $E_{\text{Fermi}}=-t$. The wave-functions $\Psi_{\uparrow , \downarrow, \lambda} (m)$ are computed from a numerical diagonalization of Eq. \eqref{simpleq}.  One clearly observes sharp peaks in the spin densities, which correspond to opposite spins at the two edges. As $\phi$ is progressively varied from  $\phi= 1$ to $\pi=2 \pi -1$ , one first observes a transition into the ``bulk" regime and finally a spin inversion at the edges (cf. Fig. 1 (c)). Let us stress that such a result is only valid for abrupt walls: in the presence of an external confining trap, the edge states and their corresponding signatures will be destroyed. In order to overcome this problem, one could induce the tunneling $t$ by laser-assisted tunneling methods, and create synthetic walls within the confined system by abruptly changing the tunneling amplitude in the central region \cite{SH}. Then, the sharp peaks illustrated in Fig. 1 (c) and corresponding to the edge states will be observed at these synthetic walls. Consequently, by varying the parameter $\phi$, and performing \emph{in situ} spin-resolved density measurements \cite{insitu}, one could directly detect the spin-pumping process illustrated in Fig. 1 (b)-(c). 

Another method would be to measure the spin Chern number SChN, which could also be evaluated from density measurements. This method is based on the fact that the Chern number of individual spin species, $\text{ChN}_{\uparrow, \downarrow}$, can be computed from the density through the Streda formula  \cite{DP}
\begin{equation}
\text{ChN}_{\uparrow}(E_{\text{Fermi}})=\frac{\Delta \rho_{\uparrow}}{\Delta \beta},\label{streda}
\end{equation}
Here, the (local) Fermi energy is supposed to lie in a bulk gap, which is associated to a plateau in the density profiles, and $\Delta \beta=\beta - \beta '$, $\Delta \rho_{\uparrow}=\rho_{\uparrow}(\beta)-\rho_{\uparrow}(\beta ') $. This equation expresses the fact that the Chern number associated to a bulk gap can be evaluated by comparing the density plateaus obtained from two configurations of the system (i.e. with $\beta$ and $\beta '$). In Fig. 3, one shows the density profiles for $\beta=1/3$ and $\beta '=1/4$, in a system confined by a harmonic potential $V_{\text{trap}}(m)=V_{\text{conf}} \times \, (m - c)^2$, where $c=L/2$. Considering a local-density approximation, one defines the Fermi energy locally as $E_{\text{fermi}}(m)=E_{\text{fermi}}^0-V_{\text{trap}}(m)$: in this regime, where the confining potential is considered to vary smoothly, the density profiles depict several plateaus that correspond to the bulk gaps located below the chemical potential $E_{\text{fermi}}^0$. In Fig. 3, the Fermi energy is set at $E_{\text{fermi}}^0=0$, thus the density plateaus correspond to the lowest bulk gap of Fig. 1 (a). The formula \eqref{streda} yields $\text{ChN}_{\uparrow}(\text{1st gap})=1$, and therefore $\nu=1$ as already evaluated above from Eq. \eqref{chern}. We note that this result is independent of the parameter $\phi$, which highlights the fact that the topological invariants (i.e. $\text{ChN}_{\uparrow, \downarrow}$, SChN and $\nu$) can indeed be defined at each value of the parameter $\phi$ (cf. discussion above and Ref. \cite{TP}).  Consequently, a density measurement at $\phi$ fixed allows to directly determine the $Z_2$ class of our 1D system in the presence of an external confining trap. Finally, we note that this detection scheme is also suited for finite spin-mixing perturbations, i.e. $\gamma \ne 0$, as long as the bulk gaps remain open.

\begin{figure}[tbp]
\includegraphics[width=8cm]{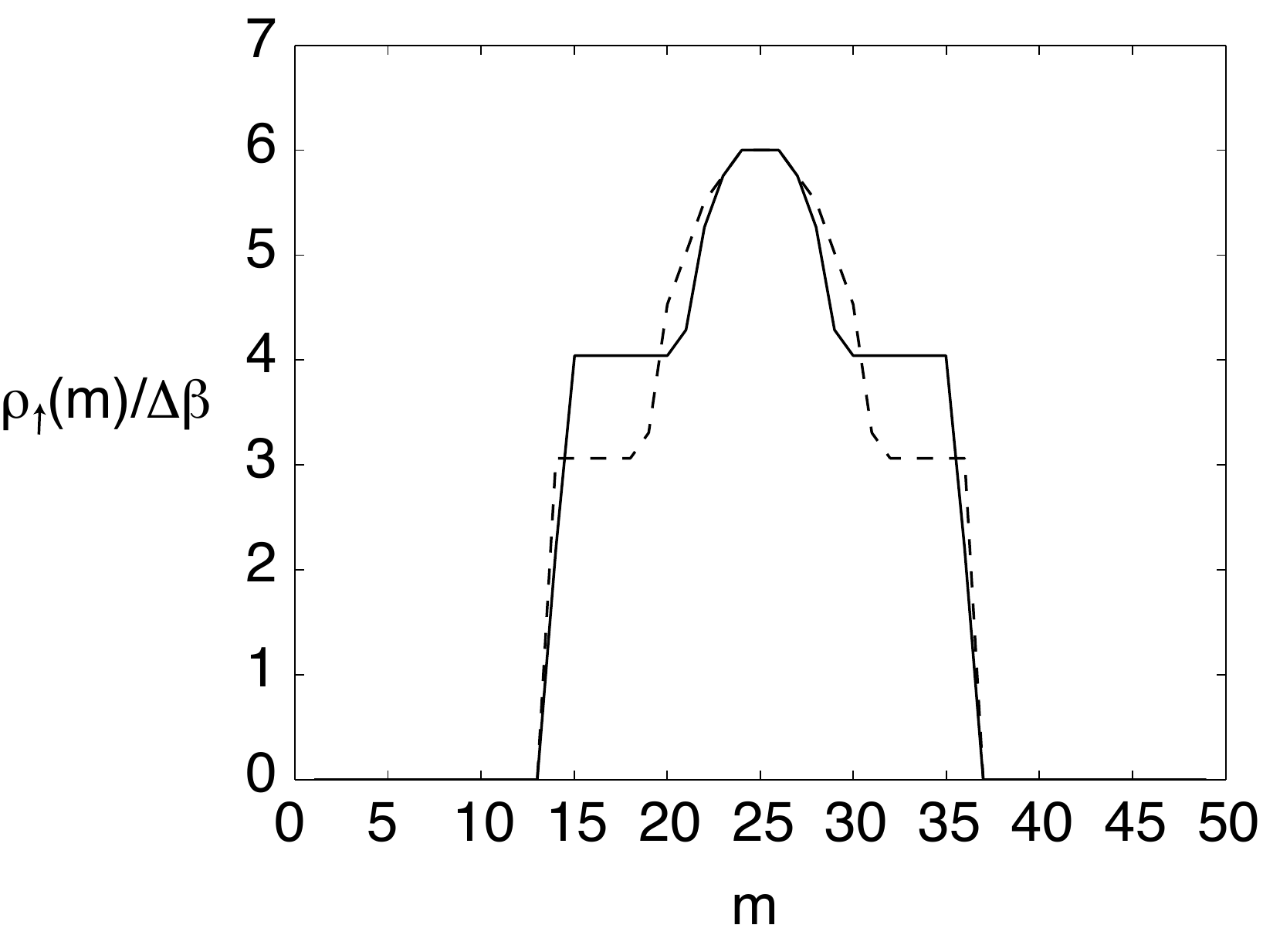}
\caption{Spin-up density $\rho_{\uparrow}(m)/\Delta \beta$ for a 1D lattice with $L=50$ sites. Here $\beta=1/3$ (blue line) and $\beta '=1/4$ (blue dotted line), $\Delta \beta= \beta - \beta ' $, the Fermi energy $E_{\text{Fermi}}^0=0$ and $\phi=\pi/2$. The system is confined by a harmonic trap $V_{\text{conf}} \, (m - c)^2$, with $V_{\text{conf}}=0.02 t$. Comparing the two plateaus indicates that $\text{ChN}_{\uparrow}=1$ inside the first bulk gap (Eq. \eqref{streda}): this density measurement attributes a non-trivial $Z_2$ index $\nu=1$ to the lowest bulk gap illustrated in Fig. 1 (a).}
\end{figure}

It was shown in Ref. {\color{blue}\cite{SH}}, that the combination of a spin-mixing perturbation $\gamma \ne 0$ and an additional staggered potential $\lambda_{\text{stag}} \ne 0$ leads to interesting phase transitions between trivial and non-trivial $Z_2$ phases. These transitions occur individually in the different bulk gaps and can be obtained by solving the Harper Eq. \eqref{harper}. Exploring these topological phase transitions with our 1D model requires to engineer the Peierls operator $\theta_x$ as well as the staggered potential, which both act along the $x$ direction (i.e. the direction of our 1D system). The realization of the hopping operator $\theta_x$ demands to control the tunneling in a spin-dependent manner, which can be achieved with several Raman transitions that act individually on the two atomic internal states (cf. Refs. \cite{GF4,SH}). On the other hand, the staggered potential could be easily produced by a weak lattice $V_{\text{stag}} \sim t$, with wave number $k_{\text{stag}}=k_1/2$. Using this configuration, one could directly detect  topological phase transitions by varying the staggered potential strength and performing spin-resolved density measurements, since the latter provide sufficient informations to classify our system in terms of the $Z_2$ index $\nu$ (cf. above). For example, the trivial phase $\nu=0$ corresponding to $\beta=1/6$, $E_{\text{Fermi}}=-t$ and $\lambda_{\text{stag}}=0$ (cf. Fig. 2) will evolve into a QSH phase with $\nu=1$ for $\lambda_{\text{stag}} > 1.5 t$ (cf. Ref. \cite{SH}).

In summary, we have proposed an experiment scheme to simulate $Z_2$ topological phases with cold atoms trapped in a 1D bichromatic optical lattice. Our scheme is based on the dimensional reduction of a 2D model exhibiting $Z_2$ topological insulating phases, and which captures its essential properties. Our 1D atomic system is described by a generalized Harper equation, which can be simulated by a spin-1/2 generalization of the Aubry-Andre system recently realized with cold atoms \cite{AL}. The latter can be practically engineered using the current technology offered by optical lattices, exploiting the interferences of a primary lattice with weak state-dependent lattices. Interestingly, our simple scheme is able to transfer spin-resolved edge states, with opposite spin components, from one edge to the other. This manipulation, which can be easily performed by varying the secondary lattices configuration \cite{spindep3}, constitutes an elegant manner for transporting topologically protected quantum states \cite{QS}. Besides, we have discussed the possibility to drive topological phase transitions, by varying an additional staggered potential.  Furthermore, we have shown that the spin Chern number, which allows to classify the topological phases of our 1D system, can be
evaluated using spin-resolved atomic density measurements. In the presence of sharp boundaries, we have shown that spin-resolved
density profiles would already present clear signatures of edge states, with opposite spin components at the two edges. Therefore, the $Z_2$ phases emanating from our 1D system could be probed with the current technologies, such as \textit{in situ} imaging techniques. Let us mention that additional signatures could be obtained through cyclotron-Bloch dynamics \cite{kolo}. We note that state-dependent lattices generally lead to large spontaneous emission rates for fermionic species, a drawback which could be avoided by considering an atom-chip realization of our model \cite{SH}.  Finally, we stress that our dimensional reduction approach may be generalized to explore three-dimensional topological phases \cite{threed} using 2D optical lattice.

The authors thank J.K. Pachos and A. Bermudez for many helpful discussions and
comments. NG thanks FRS-FNRS for financial support.
This work was supported by the NSFC (Nos. 60978009, 10974059,  and
11125417), the Major Research Plan of the NSFC ( No. 91121023),
the SKPBR of China (Nos.2009CB929604?2011CB922104 and 2011CBA00200), and NUS Academic Research (No.WBS:R-710-000-008-271).

\end{document}